\documentstyle[11pt]{article}
\begin{document}
\title{Some discussions on strong interaction}
\author{{Ning Wu}\\
{\small CCAST (World Lab), P.O.Box 8730, Beijing 100080, P.R.China}\\
{\small and}\\ 
{\small Division 1, Institute of High Energy Physics, P.O.Box 918-1, 
Beijing 100039, P.R.China}}
\maketitle
\vskip 0.8in

\vskip 0.8in

\noindent
[Abstract]  Using the gauge field model with massive gauge bosons, we could construct a new 
model to describe the strong interaction. In this new quantum chromodynamics(QCD) model, we 
will 
introduce two sets of gluon fields, one set is massive and another set is massless. Correspondingly, 
there may exists three sets of glueballs which have the same spin-parity but have different masses. 
A possible new kind of long-range force field is also discussed in this paper.
\\

~~~~ Now, it is generally believed that the most elementary constituents of matter are elementary 
fermions, such as quarks and leptons, and elementary bosons, such as photon, gluon and gauge 
bosons $W^{\pm}$ and $Z$ etc.. Quarks are elementary constituents of hadrons\lbrack 1-2 \rbrack. In 
order to solve the problem of Fermi statistics of quarks inside hadrons, the concept of color of 
quarks 
is introduced to physics\lbrack 3-5 \rbrack. It is known that quarks' dynamics obeys $SU(3)_c$ 
symmetry. Using Yang-Mills gauge field theory\lbrack 6 \rbrack, physicists have found a quantum 
theory to describe strong interaction---quantum chromodynamics (QCD). We know that, in QCD, 
there exists only one kind of gauge field. Because QCD has strict $SU(3)_c$ symmetry, the mass of 
gauge field is zero. \\

~~~~Using the gauge field model which has a set of massive gauge bosons as well as a set of 
massless gauge bosons \lbrack 7 \rbrack, we could construct another QCD to describe strong 
interaction. Because of color confinement, colored gluons can not existin nature in a state of free 
particle, they exist inside glueballs or hybrid states. Because there exist two sets of gluons, there
may exist three sets 
of glueballs in mass spectrum. They are $(g_1 g_1)$, $(g_1 g_2)$ and $(g_2 g_2)$, which have the 
same spin-parity but different masses. Although all gluons are in color octet, some gluons are 
colorless. If colorless gluons are not restricted by color confinement, these colorless gluons  may 
exist in a state of free particle and there may exist a new kind of long-range force field \lbrack
8,11\rbrack. In experiment, these free gluons may mix in $\gamma$ photons.So,physicists will find that 
$\gamma$ photons take part in strong interaction \lbrack 10 \rbrack. The existence of these free 
massless gluons may explain the phenomenon that $\gamma$ photons take part in strong interaction.
\\

~~~~ The quark fields is denoted as
$$
\psi_a (x) ~~,~~ (a=1,2,3) ,
\eqno{(1)} 
$$
where $a$ is color index and the flavor index is omitted. Define
$$
\psi  =\left ( 
\begin{array}{c}
\psi_1 \\
\psi_2 \\
\psi_3
\end{array}
\right ) . 
 \eqno{(2)} 
$$
All $\psi$ form the fundamental representative space of $SU(3)_c$. In the present theory, two 
different gauge fields $A_{1 \mu} (x)$ and $A_{2 \mu} (x)$ will be introduced. The lagrangian of 
the model is:
$$
\begin{array}{ccl}
\cal L &= &- \overline{\psi}
\lbrack \gamma ^{\mu} ( \partial _{\mu} - i g A_{1 \mu})  +m \rbrack  \psi \\ 
&&-\frac{1}{4K} Tr( A_1^{\mu \nu} A_{1 \mu \nu} )
-\frac{1}{4K} Tr( A_2^{\mu \nu} A_{2 \mu \nu} ) \\
&&-\frac{\mu ^2}{2K} 
Tr \left \lbrack ( {\rm cos} \alpha A_1^{\mu}+{\rm sin}\alpha A_2^{\mu})
( {\rm cos} \alpha A_{1 \mu}+ {\rm sin}\alpha A_{2 \mu} ) \right \rbrack
\end{array}
\eqno{(3)} 
$$
where
$$
\begin{array}{l}
A_{1 \mu \nu} = \partial _{\mu} A_{1 \nu}  - \partial _{\nu} A_{1 \mu} 
- i g \lbrack A_{1 \mu} ~ ,~   A_{1 \nu}  \rbrack  \\
A_{2 \mu \nu} = \partial _{\mu} A_{2 \nu}  - \partial _{\nu} A_{2 \mu} 
+ i g {\rm tg} \alpha \lbrack A_{2 \mu} ~ ,~   A_{2 \nu}  \rbrack  
\end{array}
\eqno{(4)} 
$$
The above lagrangian can be proved to have strict $SU(3)_c$ gauge symmetry. \\

~~~~ In the above lagrangian, $A_{1 \mu}$ and $A_{2 \mu}$ are not eigenvectors of mass matrix. 
So, let's make the following transformations:
$$
G_{ 1 \mu}= {\rm cos}\alpha  A_{1 \mu}+{\rm sin}\alpha A_{2 \mu}
\eqno{(5)} 
$$
$$
G_{ 2 \mu}= - {\rm sin}\alpha  A_{1 \mu}+{\rm cos}\alpha A_{2 \mu}
\eqno{(6)} 
$$
Then, the lagrangian given by eq(3) will change into
$$
\begin{array}{ccl}
\cal L &= &- \overline{\psi}
\lbrack \gamma ^{\mu} ( \partial _{\mu} - i g {\rm cos} \alpha G_{1 \mu}
+ i g {\rm sin} \alpha G_{2 \mu})  +m \rbrack  \psi \\ 
&&-\frac{1}{4} G_{1 0}^{i \mu \nu} G^i_{1 0 \mu \nu} 
-\frac{1}{4} G_{2 0}^{i \mu \nu} G^i_{2 0 \mu \nu} 
-\frac{\mu ^2}{2} G_1^{i \mu} G^i_{1 \mu} + {\cal L}_{g I} ,
\end{array}
\eqno{(7)} 
$$
where
$$
G^i_{m 0 \mu \nu} = \partial _{\mu} G^i_{m \nu}  - \partial _{\nu} G^i_{m \mu} 
~~~(m=1,2)
\eqno{(8)} 
$$
$$
G_{m  \mu } = G^i_{m \mu} \lambda_i /2 ,
\eqno{(9)} 
$$
and ${\cal L}_{g I}$ only contains interaction terms of gauge fields. From the above lagrangian, we 
see that, gluon fields $G_1$ are massive fields whose masses are $\mu$, and gluon fields $G_2$ are 
massless fields. We use $g_1$ and $g_2$ denote gluons correspond  to the gluon fields $G_1$ and 
$G_2$ respectively. Because of color confinement, all colored gluons must exist in bound states. 
Two kinds of gluons may form three kinds of glueballs which are denoted by $(g_1 g_1)$, $(g_1 
g_2)$ and $(g_2 g_2)$. They have the same spin-parity but different masses. Because gluon $g_1 $ 
is massive and gluon $g_2$ is massless, glueball $(g_1 g_1)$ is heavier than glueball $(g_1 g_2)$ 
and glueball $(g_1 g_2)$ is heavier than glueball $(g_2 g_2)$. If ${\rm tg} \alpha$ is not very small 
or very big, then we will see these three kinds of glueballs in the mass spectrum simultaneously. If 
${\rm tg} \alpha$ is very small, then only the peak corresponds to glueball $(g_1 g_1)$ is obvious, 
and those peaks correspond to glueballs $(g_1 g_2)$ and $(g_2 g_2)$ are very weak. In this case, 
only glueball $(g_1 g_1)$ can be seen in the mass spectrum. Similarly, if ${\rm tg} \alpha$ is very 
big, only glueball $(g_2 g_2)$ can be seen in the mass spectrum. \\

~~~~ From eq(7), we know that the coupling constant between massive gluon fields and quark is $- g 
{\rm cos} \alpha$, and the coupling constant between massless gluon fields and quark is $g {\rm sin} 
\alpha$. Because strong interaction is a short range interaction, we anticipate that the coupling 
between massive gluon fields and quark is much stronger than the coupling between massless gluon 
fields and quark. Therefore,
$$
{\rm tg} \alpha \ll 1. 
\eqno{(10)} 
$$
And the strong interaction mainly originate from the interchange of massive gluons. The range of 
strong force depends on the mass of gluon. Suppose that the range of strong force is about the radius 
of proton, then the mass of gluon is about $1 Gev$ . If this is true, the masses of $g_1^3$ and 
$g_1^8$ are also about $1 Gev$. In this case, we could prove that , in a proper approximation, the 
potential of the strong interaction is linear\lbrack 11 \rbrack. \\

~~~~ Because of color confinement, free quarks can not be directly seen by experiment though the 
masses of 
quarks are not very big. There are a lot of explanations of quark confinement at present. A possible 
explanation is to suppose that vacuum is an anticolor medium \lbrack 9 \rbrack . According to this 
point of view, all colored particles will be confined in bag and all particles detected by experiment 
are 
colorless particle. Now, let's discuss the problem on gluon confinement. In eq(9), $\lambda _i$ are 
Gell-Mann matrix. Substitute the expression of $\lambda _i$ into eq(9), we obtain
$$
G_{\mu}  =  \frac{1}{2}  \left ( 
\begin{array}{ccc}
G^3_{\mu } + \frac{1}{\sqrt{3}} G^8_{\mu}
& G^1_{\mu} - i G^2_{\mu}
& G^4_{\mu} - i G^5_{\mu}
 \\
G^1_{\mu} + i G^2_{\mu}
& - G^3_{\mu } + \frac{1}{\sqrt{3}} G^8_{\mu}
& G^6_{\mu} - i G^7_{\mu}
\\
G^4_{\mu} + i G^5_{\mu}
& G^6_{\mu} + i G^7_{\mu}
& - \frac{2}{\sqrt{3}} G^8_{\mu}
\end{array}
\right ) . 
 \eqno{(11)} 
$$
In the above relation, $G_{\mu}$ represents $G_{1 \mu}$ or $G_{2 \mu}$. All gluons in the non-
diagonal position are colored gluons. The interaction terms between diagonal gluons and quarks are
$$
\begin{array}{l}
- i g' \overline{\psi_1} \gamma ^{\mu} \psi_1 (G^3_{\mu } + \frac{1}{\sqrt{3}} G^8_{\mu} )
- i g' \overline{\psi_2} \gamma ^{\mu} \psi_2 (- G^3_{\mu } + \frac{1}{\sqrt{3}} G^8_{\mu} )
 \\ 
+ \frac{2}{\sqrt{3}}  i g' \overline{\psi_3} \gamma ^{\mu} \psi_3 G^8_{\mu }
\end{array}
\eqno{(12)} 
$$
where
$$
\left \lbrace
\begin{array}{ll}
g' = -g {\rm cos} \alpha
& if~ G_{\mu}~ is ~G_{1 \mu}
 \\
g' =  g {\rm sin} \alpha
& if~ G_{\mu}~ is ~G_{2 \mu}
\end{array}
\right.
\eqno{(13)} 
$$
Because $ \overline{\psi_m} \gamma ^{\mu} \psi_m $  $(m=1,2)$ are colorless, $G_{\mu}^3$ 
and 
$G_{\mu}^8$ are colorless gluons. If color confinement originate from the property that vacuum is 
an 
anticolor medium, colorless particle, such as $G_{\mu}^3$ or $G_{\mu}^8$, can exist in a state 
of free particle, and therefore can be detected by experiment. \\

~~~~ If gluons $g_1^3$ and $g_1^8$ can exist in the state of free particle, some of their properties 
are similar to those of glueballs, especially their decay modes. They are massive vector particles. 
They carry no electric charge. Because they couple to quarks and massless gluons, there are hadrons 
and massless, electric neutral vector particles in their decay products. Those massless, electric 
neutral vector particles may be regarded as $\gamma$ photon in the experiment. In other words, 
$g_1^3$ and $g_1^8$ may have "rediative" decay modes. Because $g_1^3$ and $g_1^8$ don't 
couple to leptons, there are almost no leptons in their decay products, and they will not be directly 
produced in the $e^+ + e^-$ collisions, but they will be produced in the strong decay or in the $ p + 
\overline{p}$ collisions or in the course of any other strong interaction. It is very important 
in both theory and experiment to search for these two kinds of gluons in the experiment. To verify 
whether $g_1^3$ and $g_1^8$ can exist as free particles will conduce to the understanding of the 
nature of color confinement.  \\

~~~~ For gluons $g_2^3$ and $g_2^8$, they are massless, electric neutral vector particle, and can 
be emitted from or be absorbed by general atoms. These properties are the same as those of photon, 
so 
in the experiment, they may be regarded as photon. But they have essential differences. It is known 
that $g_2^3$ and $g_2^8$ take part in strong interaction, and photon does not directly take part in 
strong interaction. If physicists find that photons take part in strong interaction obviously, that means 
that there are massless, electric neutral vector particles mixed in photons. In other words, those 
photons take part in strong interaction 
might be $g_2^3$ or $g_2^8$ gluons. Indeed, physicists have found that photons take part in strong 
interaction \lbrack 10 \rbrack, it is important to verify whether there are gluons mixed in 
those photons. Another important difference between $g_2^3$ or $g_2^8$ and photon is that 
$g_2^3$ and $g_2^8$ don't interact with charged leptons $e, \mu$ and $\tau$, but photons interact 
with any charged particles include leptons. Finally, gluons have self interaction, but photons have no 
self interaction. From theoretical point of view, photon is a particle of no structure, it should not have 
structure functions. From the present experimental results \lbrack 10 \rbrack, some gluons might 
mixed in photons, and the so-called structure functions of photon might be those of gluons. \\

~~~~ There may exist a kind of long-range force field transmit by massless gluons $g_2^3$ and 
$g_2^8$. Correspondingly, there may exists a kind of wave transmit by those gluons. Because all 
colored particles are confined and all matter exists in nature is colorless, the influences of this long-
range force field to the motion of matter is very very small.

\section*{Reference:}
\begin{description}
\item[\lbrack 1 \rbrack]  M.Gell-Mann, Phys.Lett. {\bf 8} (1964) 214
\item[\lbrack 2 \rbrack]  G.Zweig, CERN preprints, Th. 401 and 412, (1964)
\item[\lbrack 3 \rbrack]  M.Han, Y.Nambu, Phys.Rev. {\bf B139} (1965) 1006
\item[\lbrack 4 \rbrack]  H.Fritsch, M.Gell-Mann, Proc.XVI Int. Conf. on High Energy Phys., 
Vol2, 
P.135 (Chicago)
\item[\lbrack 5 \rbrack]  H.Fritsch, M.Gell-Mann, H.Leutwyler, Phys.Lett {\bf B97} (1980)365
\item[\lbrack 6 \rbrack]  C.N.Yang, R.L.Mills, Phys Rev {\bf 96} (1954) 191
\item[\lbrack 7 \rbrack]  Ning Wu, Gauge Field Model With Massive Gauge Bosons, hep-
ph/9802236
\item[\lbrack 8 \rbrack]  Ning Wu, A New Electroweak Model without Higgs
particle, hep-ph/9802237
\item[\lbrack 9 \rbrack]  T.D.Lee, Particle Physics and Introduction to Field Theory, (Harwood 
Academic Publishers, Amsterdam, 1983)
\item[\lbrack 10 \rbrack] There are many papers discuss the photon interaction experimentally. See 
for example hep-ex /9711005
\item[\lbrack 11 \rbrack] Ning Wu,(in preparation)
\end{description}

\end{document}